\documentstyle[sprocl]{article}

\def\prb{Phys. Rev. B}
\def\prl{Phys. Rev. Lett.}

\def\be{\begin{equation}}
\def\ee{\end{equation}}
\def\ba{\begin{eqnarray}}
\def\ea{\end{eqnarray}}
\def\LSCO{La$_{2-x}$Sr$_x$CuO$_4$}
\def\BSCCO{Bi$_2$Sr$_2$CaCu$_2$O$_{8+\delta}$}
\def\C60{A$_x$C$_{60}$}

\def\hts{high temperature superconductors}
\def\htr{high temperature superconductor}
\def\qp{quasiparticle}
\def\qps{quasiparticles}
\def\qcp{quantum critical point}
\begin{document}

\title{CHARGE TRANSPORT IN SYNTHETIC METALS}

\author{V.~J.~Emery$^1$, S.~A.~Kivelson$^2$, and V.~N.~Muthukumar$^1$}
\address{
$^1$Dept. of Physics
Brookhaven National Laboratory
Upton, NY  11973-5000}
\address{
$^2$Dept. of Physics
University of California at Los Angeles
Los Angeles, CA 90095}

\maketitle\abstracts
{The phenomenology of charge transport in synthetic metals is 
reviewed. It is argued that the conventional {\qp} picture and
Boltzmann transport theory do not apply to these materials. The central
ideas of Fermi liquid theory are reviewed, and the significant corrections
produced by {\qp} scattering from ferromagnetic spin fluctuations in liquid
$^3$He are described. It is shown that Sr$_2$RuO$_4$ does not display the 
symptoms of a nearly-ferromagnetic Fermi liquid, so the source of its 
odd angular momentum pairing remains to be understood.
The solution of an assisted-tunneling model of charge transport in
quasi-one dimensional materials is described. This model has a quantum 
critical point and gives a resistivity that is linear in temperature or 
frequency, whichever is greater.}




\section{The Nature of the Problem}

The twentieth century has seen the development of an outstandingly successful
theory of the solid state, including a theory of the equilibrium and
transport properties of metals.\cite{history} For transport properties,
the central concept of a mean free path $l$ was introduced at the beginning
of the century.\cite{mfp,loren} However a number of interesting synthetic 
metals discovered in the past few decades seem to violate  the conventional 
theory. They are ``bad metals'' \cite{badmetal} in the sense that their 
resistivity $\rho$ has a metallic temperature dependence (it increases with 
the temperature $T$) but, if $\rho$ were interpreted in terms of a mean free 
path, the uncertainty $2\pi/l$ in the Fermi wave vector $k_F$ would be greater 
than $k_F$ itself, so the concept of an electron state in momentum space would 
be ill defined. This is known as the Ioffe-Regel condition.\cite{IR} 
Among the materials in question are the {\hts}; other oxides including the 
ruthenates, such as SrRuO$_3$ and Sr$_2$RuO$_4$; organic conductors, such as 
TTF-TCNQ; and alkali-doped $C_{60}$. 

In all of these materials, $\rho$ decreases as $T$ decreases and, 
unless a phase transition intervenes, $\rho$ could become small enough to 
give the appearance that the conventional theory is applicable. However, 
for many synthetic metals, $\rho$ is linear in $T$ over an extended range of
temperatures, and it would be inconsistent to explain one and the same 
phenomenon by the conventional theory at low temperatures, and by some other 
theory at high temperatures. 

Another issue is the role of phonons. Typically the $T$-linear resistivity
extends from well above to well below phonon Debye temperatures, so it cannot 
be attributed to scattering from classical phonons. But there is no change in 
the slope of $\rho$, so it appears that the phonons make no contribution at all 
to the resistivity, even in materials in which there is independent evidence 
of a significant electron-phonon coupling. This behavior is quite inconsistent 
with the usual Boltzmann transport theory in which scattering rates are 
additive.

Alkali-doped $C_{60}$ is a particularly interesting example of a bad metal
because $\rho$ is $T$-linear but, in a {\qp} picture, it would 
correspond to a mean free path that is less than the size of a $C_{60}$ 
molecule, and can even be as small as a carbon-carbon distance,\cite{hellman} 
which is less than 10\% of the size of a unit cell. Then it clearly does not 
make sense to interpret transport in terms of {\qps} freely propagating from 
cell to cell.

TTF-TCNQ is an organic conductor consisting of separate linear stacks of TTF 
molecules and TCNQ molecules. For a substantial range of temperatures, the 
molecular stacks are weakly coupled and the material behaves as an array 
of one-dimensional electron gases, for which the scattering rate of an
electron from an impurity is proportional to $T^{-\alpha}$,
where $\alpha$ is positive if charge-density wave (CDW) correlations are
enhanced.\cite{lp,mattis} In fact TTF-TCNQ does undergo a CDW transition at 
low temperatures, so the resistivity should decrease as the temperature is
increased. However, at constant inter-molecular spacing along a stack,
TTF-TCNQ has a resistivity that {\it increases} linearly with T in
the one-dimensional region,\cite{cooper} where it also violates
the Ioffe-Regel condition. Thus there is not even a qualitative understanding 
of charge transport in this material.

These examples illustrate the point that the transport properties of 
synthetic metals pose a problem of fundamental importance for condensed
matter physics, inasmuch as the conventional transport theory is inapplicable. 
Other examples will be given in the next section. One possible solution
is that transport is controlled by a {\qcp}, {\it i.e.} a critical point at 
zero temperature. This could be a zero-dimensional {\qcp}, as in the physics 
of the two-channel Kondo problem,\cite{kondo} or a two or three dimensional 
{\qcp}, generated by long-distance physics.\cite{varma,castellani} More work
is needed to establish whether this idea provides a general solution of the
problem, or if other new paradigms will have to be developed. However, at a 
{\qcp}, the electron-phonon coupling could be irrelevant, so there is the 
possibility of explaining the lack of a phonon contribution to $\rho$. A 
zero-dimensional quantum critical model of charge transport via assisted 
tunnelling will be described in Sec. III.

In principle, angular resolved photoemission spectroscopy is capable of
verifying or disproving the {\qp} picture, provided the resolution is
sufficiently good. So far the best such experiments have been carried
out on the {\htr} {\BSCCO} and, typically,  they do not validate the 
existence of {\qps} in the normal state.\cite{arpes} Now, a new generation of 
spectrometers with much improved resolution is at hand and, in the future, we 
may have better input from this spectroscopy.

\section{Conventional Transport Theory}

\subsection{Drude-Lorentz Theory}

A simplified view of the central idea of the Drude-Lorentz theory of 
transport in metals is as follows. Consider a set of electrons of density 
$n$, charge $e$ and mass $m$ moving with velocity $v$. The current density 
is $j=nev$. Now, if an electron experiences a constant electric field $E$  
during a mean free time $\tau$, the velocity is $eE\tau/m$, and the 
conductivity $\sigma_0$, defined from $j=\sigma_0 E$, is given by 
\be
\sigma_0 = {ne^2 \over m} \tau
\ee
where $ne^2/m$ is the ``Drude weight''. This simple derivation emphasises
the central assumption of an electron that is freely-moving during a mean 
free time $\tau$. Of course, in practice, all of the work goes into 
calculating the mean free time from electron-electron collisions and from 
the scattering of electrons by phonons and impurities. After the
invention of quantum mechanics, the practical implementation of these 
calculations changed drastically.

An estimate of the Ioffe-Regel value of $\rho$ gives about 
0.2 to 0.3 m$\Omega$cm. Here, it was assumed that, in three dimensions, 
$k_F^{-1}$ is about 1\AA. In two dimensions, the value of $k_F$ does not 
enter into the estimate in because $n \sim k_F^2$ and 
\be
\hbar k_F^2 \tau/m =  k_F v_F \tau =k_F l
\ee
Thus the magnitude of $k_F l$ is independent of issues, such as the existence 
of a large or small Fermi surface. In two dimensions, the length scale that
enters into $\rho$ is the spacing between planes, for which we have used 6\AA, 
a common value in the {\hts}, which are prime examples of bad metals. In 
general there also should be a geometric factor that expresses the shape of 
the ``Fermi surface.''

This estimate of the Ioffe-Regel condition is widely violated.
In optimally doped {\LSCO}, the resistivity in the a-b plane increases linearly 
with temperature from $T_c$ up to 900$^{\circ}$K, where its magnitude is 
0.7 m$\Omega$cm.\cite{takagi}  Other high temperature superconductors have 
similarly high resistivities at elevated temperatures.
In Rb$_3$C$_{60}$, the resistivity has been measured\cite{HC60} up to 520K, 
where the implied mean free path is $0.6\AA$, corresponding to a value of 
$k_F l/2\pi =0.03$!  In TTF-TCNQ, the room-temperature resistivity in the 
highly conducting direction is about 2m$\Omega$-cm.\cite{cooper}

\subsection{Resistivity Saturation.} 

In normal metals with strong electron-phonon coupling, the resistivity is 
well described by Boltzmann transport theory at low temperatures. At high 
temperatures the resistivity does not violate the Ioffe-Regel condition, but
saturates before doing so. For example, the resistivity of Nb$_3$Sn saturates
at a value of about 0.1m$\Omega-cm$.

\subsection{Fermi Liquid Theory}

In the mid fifties, Landau introduced the concept of a Fermi liquid, in which 
the behavior of the system is described in terms of elementary excitations out
of the ground state.\cite{landau} The excitations were assumed to be fermionic 
{\qps} and bosonic collective excitations, and the idea was that, even in a 
strongly-correlated system, there are few such excitations at low temperatures, 
so transport coefficients can be evaluated by a quantum Boltzmann transport
theory which includes two-particle collision processes. 
In this way, it is straightforward to calculate the general temperature 
and frequency dependence of transport and equilibrium properties, but the 
system-dependent {\it magnitudes} of these quantities are specified in terms of a 
set of phenomenological Fermi liquid parameters. This theory was devised for 
liquid $^3$He, but it soon was applied to metals and to atomic nuclei. 

A major consequence of the theory is that
scattering rates $\tau^{-1}$ are proportional to $T^2$ at low temperatures.
However, liquid $^3$He is a nearly-ferromagnetic Fermi liquid, and the
scattering of quasiparticles from spin fluctuations gives a significant $T^3$
correction to the scattering rate:\cite{nearfm}
\be
{1 \over \tau} = aT^2-bT^3
\ee
where $b/a$ is large enough\cite{batsheva} to make the $T^3$ term a 25\% 
correction at a temperature of 40mK. The scattering rate is never simply
proportional to $T^2$, and indeed, for some time, this behavior confounded 
attempts to verify the predictions of Fermi liquid theory in liquid $^3$He.
Another consequence of the nearly ferromagnetic nature of liquid $^3$He is that
it is a fermion superfluid with pairing in odd angular momentum 
states.\cite{odd} 

It is  particularly interesting to consider Sr$_2$RuO$_4$ in the same context.
This material has a $T$-linear resistivity\cite{forro} up to 1100K but clean 
samples cross over to the Fermi liquid $T^2$ behavior below 30K.\cite{maeno}  
Sr$_2$RuO$_4$ is a superconductor with a transition temperature of about 1K. 
It has been noted that this material should have a strong Hund's rule coupling,
and that a related material SrRuO$_3$ has a ferromagnetic 
ground state.\cite{sigrice} This lead 
to the suggestion that Sr$_2$RuO$_4$ may be a nearly-ferromagnetic Fermi 
liquid at low temperatures,\cite{sigrice} and to the (apparently successful) 
prediction that Sr$_2$RuO$_4$, like $^3$He, should have pairing in odd angular 
momentum states. However, it is not so clear that Sr$_2$RuO$_4$ really is
nearly ferromagnetic: its resistivity is proportional to $T^2$ up to 
30K \cite{andymc}, and shows no sign of the $T^3$ contribution that would
characterize the scattering of electrons from ferromagnetic spin 
fluctuations. It may well be that new physics is involved, and that the 
odd-angular momentum pairing in Sr$_2$RuO$_4$ evolves from the 
crossover from a bad metal to a Fermi liquid as the temperature is reduced.

\section{Charge Transport in Bad metals}

\subsection{Common Features of Bad Metals}

There are a number of common features of bad metals,
although they are not universal. They are often strongly-correlated 
electron systems that are close to a Mott metal-insulator transition 
and exhibit poor screening. Frequently they are charge-transfer materials
with a low carrier concentration and show charge-density or spin-density
wave order or fluctuations. They often exhibit charge inhomogeneity, either 
self-organized (as in the {\hts} \cite{stripe}) or in virtue of their 
structure (as in alkali-doped C$_{60}$ or organic conductors).
Their resistivity is often linear in temperature
$T$ or frequency $\omega$ (whichever is greater). The property that there is 
no energy scale for $T$ and $\omega$ is characteristic 
of a quantum critical point. 

\subsection{A Model for Quasi-One Dimensional Materials}

In the absence of fermionic {\qps}, the theoretical challenge is to 
understand the nature of the processes which contribute to the low 
frequency response. This section describes one possibility -- assisted
tunnelling.

If there is no hopping transverse to the organic stacks in a quasi-one 
dimensional material, such as TTF-TCNQ, there is a flat Fermi surface.
Introduction of a small transverse hopping amplitude $t_{\perp}$ 
will cause a modulation of the Fermi surface. However if $t_{\perp} < \pi T$,
this modulation is smaller than the thermal broadening of the Fermi surface
and there is no coherent motion of the charges from one chain to another.
This is the definition of the one-dimensional region. Interactions actually
renormalize the value of $t_{\perp}$ downwards,\cite{claude} and expand 
the one-dimensional region of temperatures. However, despite the absence of 
coherent tunnelling of electrons in plane wave states, there remain 
{\it local} hopping processes between neighboring stacks, and they enable an 
electron to circumvent an impurity, which would otherwise impede its motion
along a given organic stack. These processes take advantage of the structural 
inhomogeneity and allow an electron to tunnel around the impurity. 
They dominate the conductivity because otherwise
the central stack would become insulating at low temperature. However 
this is not ordinary tunnelling; as we shall see it is coupled to the 
low-energy electronic degrees of freedom on the adjacent stacks.

Suppose the stacks are horizontal, and that there is an impurity at 
the origin on a given stack. Now introduce a pseudospin variable
${\vec \tau}$ {\it for that stack} such that $\tau^z = \pm 1$ designates the 
presence of an electron tunnelling to the right ($\tau^z = + 1$) or left 
($\tau^z = - 1$) across the impurity.  {\it For an adjacent stack}, there are 
electron variables $\psi_{\sigma,R}(x)$ and $\psi_{\sigma,L}(x)$ that annihilate 
respectively right-going and left-going electrons with spin $\sigma$ at 
position $x$. The Hamiltonian is 
\be
H = H_0 + H_{\parallel} + H_{\perp}
\ee
where 
\be
H_0 = iv_F \sum_{\sigma}\int dx 
\big[ \psi^{\dagger}_{\sigma,R}\partial_x \psi_{\sigma,R} -
\psi^{\dagger}_{\sigma,L}\partial_x \psi_{\sigma,L} \big],
\ee
\be
H_{\parallel} =  J_{\parallel} \tau^z \sum_{\sigma}
\big[ \psi^{\dagger}_{\sigma,R}(0) \psi_{\sigma,R}(0) -
\psi^{\dagger}_{\sigma,L}(0) \psi_{\sigma,L}(0) \big],
\ee
and
\be
H_{\perp} = J_{\perp}  \sum_{\sigma}
\big[ \tau^+ \psi^{\dagger}_{\sigma,R}(0) \psi_{\sigma,L}(0) +H.c.\big].
\ee
Here, $H_0$ is the kinetic energy, $H_{\parallel}$ is a current-current 
coupling, and $H_{\perp}$ is a backscattering interaction. It can be shown
that the same model is relevant for assisted tunnelling involving several 
adjacent stacks. 
                           
This problem may be solved in the same way as the two-channel Kondo
problem.\cite{orbital} Here the pseudospin ${\vec \tau}$ plays the role
of the Kondo spin and $\sigma$ plays the role of a flavor label. 
It has been shown \cite{orbital,2ck} that the 
frequency-dependent conductivity at this point is given by
\be
\sigma({\omega}) = const. \hskip 0.1cm {tanh ({\omega \over 2 T}) \over \omega}
\hskip 0.2cm {\Gamma \over \omega^2 + \Gamma^2}
\ee
From this equation it can be seen that
\ba
\sigma(\omega) \sim& {1 \over \omega}, \hskip  0.5cm for \hskip 0.2cm
T< \omega << \Gamma \nonumber\\
               \sim& {1 \over T},  \hskip 0.5cm   for \hskip 0.2cm 
\omega < T << \Gamma 
\ea
Consequently, apart from the (large) cutoff $\Gamma$, T sets the scale
for $\omega$ and $\omega$ sets the scale for T. This behavior is a 
consequence of the fact that the two-channel Kondo problem is governed
by a (0+1)-dimensional quantum critical point. ({\it i.e.} a quantum critical 
point in zero space dimensions and one time dimension.) Clearly, this
expression gives the linear temperature dependence of the d.c. resistivity
and it also implies a frequency-dependent conductivity that is proportional 
to $\omega$, provided $\omega > T$.
This behavior will be obtained whenever there is local tunnelling assisted
by low-energy degrees of freedom. In the case of alkali-doped C$_{60}$, it is
unlikely that there are sufficient low-energy {\it electronic} degrees of
freedom to give this behavior, but it could come from coupling to phonon 
modes. Other quantum critical Hamiltonians (with higher space dimension) have 
been proposed\cite{varma,castellani} to account for 
charge transport in the {\hts}, but a zero-space dimensional model has the
advantage that it is more robust in disorderly materials.

A major implication of this model is that there is enhancement of composite 
pairing of electrons (which is equivalent to odd-frequency pairing) on the 
adjacent stacks in the neighborhood of the impurity.\cite{2ck} Thus, the 
$T$-linear resistivity is rather like a paraconductivity produced by 
superconducting fluctuations, although, so far, an odd-frequency pairing 
state has not been realised in practice.

So far, electron-electron interactions have not been included in the model. 
It is known\cite{logs} that they produce logarithmic corrections to the 
correlation functions of the spin degrees of freedom that are responsible for 
the frequency and temperature dependence of the conductivity.\cite{orbital}
Interestingly, it has been found that, at low temperatures, the $T$-linear 
resistivity either crosses over to a $\ln T$ behavior,\cite{Boebinger} or has 
a $\ln T$ contribution added to it.\cite{jtran} Whether this behavior can be
accounted for by adding interactions to the model is currently under
investigation.\cite{mek}

\section*{Acknowledgements}  
This work was supported at UCLA by the National Science Foundation grant 
number DMR93-12606 and, at Brookhaven,  by the Division of Materials Sciences,
U. S. Department of Energy under contract No. DE-AC02-98CH10886.


\section*{References}
\begin{thebibliography}{99}



\bibitem{history} Proceedings of the Symposium on ``The Beginnings of Solid 
State Physics'', Proc. Roy. Soc. Lond. A {\bf 371}, 1-177 (1980).

\bibitem{mfp} P.~Drude, Annalen Phys., {\bf 1}, 566 (1900); {\it ibid},
{\bf 3}, 369 (1900).

\bibitem{loren} H.~A.~Lorentz, Proc. Acad. Amst, {\bf 7}, 438, 585, 684 (1905).

\bibitem{badmetal} V.~J.~Emery and S.~A.~Kivelson, {\prl} {\bf 74}, 
3253 (1995).

\bibitem {IR}  A.~F.~Ioffe and A.~R.~Regel, Semicond. {\bf 4}, 237 (1960).

\bibitem{hellman} F.~Hellman, private communication.

\bibitem{lp} A.~H.~Luther and I.~Peschel, {\prl} {\bf 32}, 992 (1974). 

\bibitem{mattis} D.~C.~Mattis, {\prl} {\bf 32}, 714 (1974).

\bibitem{cooper} J.~R.~Cooper, {\prb} {\bf 19}, 2404 (1979).

\bibitem{kondo} V.~J.~Emery and S.~A.~Kivelson, Physica C {\bf 235-240},
189 (1994).

\bibitem{varma} C.~M.~Varma, {\prb} {\bf 55}, 14554 (1997). 

\bibitem{castellani} C.~Castellani, C.~Di Castro, and M.~Grilli,
Z. Phys. B {\bf 103}, 137 (1997).

\bibitem{arpes} Z.-X.~Shen {\it et al.}, {\it Science} {\bf 267}, 343 (1995).

\bibitem{takagi} H.~Takagi, {\it et al.}, {\prl} {\bf 69}, 2975 (1992).

\bibitem{HC60} A.~F.~Hebard {\it et al.}, {\prb} {\bf 48}, 9945 (1993):
 W.~A.~Vareka and A.~Zettl, {\prl} {\bf 72}, 4121 (1994).

\bibitem{landau} L.~D.~Landau, J. Eksp. Teor. Fiz. {\bf 30}, 1058 (1956)
[English Transl. Soviet Physics JETP, {\bf 3}, 920 (1957)];
J. Eksp. Teor. Fiz. {\bf 32}, 59 (1957); [English Transl. Soviet Physics JETP, 
{\bf 5}, 101 (1957)].

\bibitem{nearfm} V.~J.~Emery, Phys. Rev. {\bf 170}, 205 (1968). 

\bibitem{batsheva} V.~J.~Emery in {\it Quantum Fluids}, edited by
N.~Wiser and D.~J.~Amit (Gordon and Breach, New York, 1970). p. 343.

\bibitem{odd} V.~J.~Emery, Annals of Phys. {\bf 28}, 1 (1964).

\bibitem{forro} H.~Berger, L.~Forro, and D.~Pavuna, Europhys. Lett. {\bf 41},
531 (1998);
A.~W.~Tyler, A.~P.~Mackenzie, S.~Nishizaki, and Y.~Maeno, 
to be published.

\bibitem{maeno} Y.~Maeno {\it et al.}, {\it Nature} {\bf 372}, 532 (1994).

\bibitem{sigrice} T.~M.~Rice and M.~Sigrist, J. Phys. Condens. Matter {\bf 7},
L643 (1995).

\bibitem{andymc} A.~P.~Mackenzie, private communication.

\bibitem{stripe} S.~A.~Kivelson and V.~J.~Emery in {\it Strongly Correlated 
Electronic Materials:  The Los Alamos Symposium 1993}, edited by K.S. Bedell, 
Z. Wang, D.E. Meltzer, A.V. Balatsky, and E. Abrahams, (Addison-Wesley, 
Reading, Massachusetts, 1994) p. 619.

\bibitem{claude} C.~Bourbonnais in {\it Low Dimensional Conductors and
Superconductors}, edited byD.~J\'erome and L.~G.~Caron (Plenum, New York, 1987)
p. 155.

\bibitem{orbital} V.~J.~Emery and S.~A.~Kivelson, {\prl} {\bf 71}, 3701 (1993).

\bibitem{2ck} V.~J.~Emery and S.~A.~Kivelson, {\prb} {\bf 46}, 10812 (1992).


\bibitem{logs}  For a recent reference that gives logarithmic corrections to
correlation functions, see T.~Giamarchi and H.~J.~Schulz, {\prb} {\bf 39}, 4620 (1989).

\bibitem{Boebinger} G.~Boebinger, this conference.

\bibitem{jtran} J.~M.~Tranquada {\it et al.}, {\prb} {\bf 54}, 7489 (1996). 

\bibitem{mek} V.~N.~Muthukumar, V.~J.~Emery, and S.~A.~Kivelson, work in
progress.

\end{thebibliography}
\end{document}